\documentclass[a4paper,aps,prd,10pt,preprintnumbers,twocolumn,superscriptaddress,nofootinbib,amsmath,amssymb]{revtex4-1}
\usepackage{graphicx}
\usepackage[utf8]{inputenc}
\usepackage[T1]{fontenc}
\usepackage{cmap}
\usepackage{ulem}
\DeclareMathAlphabet{\pazocal}{OMS}{zplm}{m}{n}
\def\imo{i}

\def\K{{\cal K}}

\begin{document}
\title{Conformal Weyl gravity via two stages of quasinormal ringing and late-time behavior}
\author{R. A. Konoplya} \email{roman.konoplya@gmail.com}
\affiliation{Institute of Physics and Research Centre of Theoretical Physics and Astrophysics, Faculty of Philosophy and Science, Silesian University in Opava, CZ-746 01 Opava, Czech Republic}
\affiliation{Peoples Friendship University of Russia (RUDN University), 6 Miklukho-Maklaya Street, Moscow 117198, Russian Federation}
\begin{abstract}
Black hole (BH) solution in the conformal Weyl gravity is a generalization of the Schwarzschild spacetime which includes two additional constants appearing when integrating the third order differential equations for gravitational field. One constant looks like the effective cosmological constant providing the de Sitter asymptotic of the solution. The other constant allows one to describe flat rotation of galaxies without introducing of the dark matter. Here we show that the effective ``dark matter'' term in the metric function drastically changes the asymptotic behavior of the  evolution of the wave function of a scalar field: after the Schwarzschild-like ringing phase, the ringing at another, non-Schwarzschildian, longer-lived frequency takes place before the beginning of the exponential asymptotic tail. Thus the evolution of the scalar field consists of the three qualitatively different stages: the Schwarzschild-like ringing phase, the effective dark matter ringing phase and the de Sitter phase characterized by exponential tails. The late-time behavior of the electromagnetic field is qualitatively different as well: the exponential tails appear even in the absence of the effective de Sitter term.
\end{abstract}
\maketitle

\section{Introduction}

Evolution of perturbations around black holes can be conditionally divided into the three stages: initial outburst,  quasinormal ringing, which is followed by power-law (for asymptotically flat spacetimes) or exponential (for asymptotically de Sitter spacetimes) tails at late times.  Although the stage of quasinormal ringing \cite{Konoplya:2011qq,Kokkotas:1999bd,Berti:2009kk} is the most important for current observations of gravitational waves \cite{LIGO}, the asymptotic regime at late times, represented by tails, also attracted considerable interest   \cite{Price267,Bicak268,Ching266,Ching:1994bd,Burko:2004jn,Brady270,Churilova:2019qph,Konoplya:2006gq,Rogatko:2008ut,Gibbons:2008gg,Gibbons:2008rs,Rogatko:2007zz,Moderski:2001gt,275Jing,Brady:1996za,274Koyama,Konoplya:2005et,Konoplya:2013rxa}, because only by achieving the asymptotic regime one can have the full picture of the compact object's response  to perturbations and judge about stability, duration of the ring-down phase and echoes \cite{Cardoso:2017cqb}.

In a seminal work by R. Price \cite{Price267} it was shown that massless scalar and gravitational fields around the Schwarzschild black hole decay according to the power law
\begin{equation}
|\Psi| \sim t^{-(2 \ell + 3)}.
\end{equation}
For Schwarzschild-de Sitter solution the asymptotic decay of the massless field is not power law anymore, but the exponential one \cite{Brady270}
\begin{equation}
|\Psi_{s}| \sim e^{-\ell k_{c} t}, \quad \ell=1, 2, ...,
\end{equation}
\begin{equation}
|\Psi_{s}| \sim |\Psi_{0}| + |\Psi_{1}|  e^{-2 k_{c} t}, \quad \ell=0.
\end{equation}
The electromagnetic field falls off in the Schwarzschild-de Sitter background at asymptotically late times according to the exponential law as well \cite{Molina:2003dc},
\begin{equation}
|\Psi_{el}| \sim e^{-\ell k_{el} t}, \quad \ell=1, 2, ...,
\end{equation}
where $k_{el}$ is some constant.

An interesting model of gravity is  the Weyl conformal gravity \cite{Bach}, where the effective cosmological constant appears in the background solution as an integration constant \cite{MK}, that is, without introduction of the cosmological constant into action. The latter has the form
 \begin{align}
S = \int d^4x \sqrt{-g} C_{abcd} C^{abcd},
\label{Weyl-action}
\end{align}
where $g$ is the determinant of the metric. Birkoff's theorem holds in the Weyl conformal gravity as well \cite{Riegert}.
The static and spherically symmetric vacuum solution describing a black hole in this theory was obtained by Mannheim and Kazanas \cite{MK}. This solution depends on the three parameters, $\beta$, $\gamma$ and $k$, so that the metric function $B(r)$, used in the line element
\begin{align}
d s^2 = -B(r) d t^2 + B^{-1}(r) d r^2 + r^2 (d\theta^2 + \sin^2 \theta d\phi^2),
\label{ds-Weyl}
\end{align}
can be written in the following form:
\begin{align}
B(r) = 1 - 3 \beta \gamma - \frac{2 \beta- 3 \gamma \beta^2}{r} + \gamma r -k r^2 .
\end{align}

Mannheim and Kazanas argued that the Weyl gravity can explain
the flat rotation of galaxies without introducing dark matter,
for which $\gamma$ is of the order of the inverse of the Hubble radius \cite{MK}. The astrophysical relevance of this solution was further confirmed in a number of works \cite{Islam:2018ymd,Dutta:2018oaj,Christodoulou:2018xxw}. This black hole solution \cite{MK} has been recently studied in a number of papers, with the emphasis to lensing and particle motion \cite{Takizawa:2020dja,Kasikci:2018mtg,Fathi:2020sey,Turner:2020gxo,Li:2020wvn,Fathi:2019jgd,Fathi:2020sfw}, thermodynamics \cite{Lanteri:2020trb,Xu:2018liy} and
 quasinormal modes \cite{Momennia:2019cfd,Mehrab-Momennia,Momennia:2019edt,Sharif:2020icx}.

Quasinormal modes of the Mannheim-Kazanas black hole have been recently studied in \cite{Momennia:2019cfd,Mehrab-Momennia,Momennia:2019edt} with some flaws and omissions. Thus, in \cite{Momennia:2019cfd} the wave equation for a test scalar field was identified with the P\"{o}shl-Teller equation for the near extremal values of the cosmological constant and it was stated that the stability of the scalar perturbations is proved. However, as we will show here, the effective potential at the lowest multipole number $\ell=0$ has a negative gap which is deeper exactly for the near extremal regime. Therefore, the fitting to the P\"{o}shl-Teller potential is not possible in this case and, moreover, the existence of bound states with negative energy leading to possible instability must be separately studied.
Then, in \cite{Momennia:2019edt} the obtained master wave equation for gravitational perturbations cannot be accepted, because the perturbations were fulfilled not in the conformal Weyl theory, but in another theory allowing for the same family of metrics.

However, the most interesting phenomenon which was omitted in these studies is connected with the evolution of perturbations at late times. Here we will show that the decay of a signal in the Mannheim-Kazanas background at late times is qualitatively different from that in the Schwarzschild or Schwarzschild-de Sitter cases. We will show that once the effective cosmological term is zero, the late times tails of the massless scalar field are oscillatory enveloped by the universal power-law decay. When the effective cosmological constant is turned on, this oscillatory tail becomes exponential and represents quasinormal ringing dominated by a non-Schwarzschildian frequency. On the contrary, electromagnetic perturbations decay according to the exponential law even when the effective cosmological term is zero.

The paper is organized as follows. In Sec. II we briefly discusses the wave equations for the scalar and electromagnetic perturbations. Section III relates the WKB and time-domain integration methods we used as well as the main results on quasinormal modes and late-time tails. In the conclusions we summarize the obtained results and discusses the open questions.

\section{The wave equations}

The general covariant equation for a massless scalar field has the form
\begin{equation}\label{KGg}
\frac{1}{\sqrt{-g}}\partial_\mu \left(\sqrt{-g}g^{\mu \nu}\partial_\nu\Phi\right)=0,
\end{equation}
while for an electromagnetic field it can be written as follows:
\begin{equation}\label{EmagEq}
\frac{1}{\sqrt{-g}}\partial_\mu \left(F_{\rho\sigma}g^{\rho \nu}g^{\sigma \mu}\sqrt{-g}\right)=0\,,
\end{equation}
where $\mu, \nu =0, 1, 2, 3$ and $F_{\rho\sigma}=\partial_\rho A_{\sigma}-\partial_\sigma A_{\rho}$ and $A_\mu$ is a vector potential.

After some algebra one can separate the angular variables in Eqs. (\ref{KGg},\ref{EmagEq}) and rewrite the wave equations in the following general master form
\begin{equation}  \label{klein-Gordon}
\dfrac{d^2 \Psi}{dr_*^2}+(\omega^2-V(r))\Psi=0,
\end{equation}
in terms of the ``tortoise coordinate'' $r_*$ \cite{Konoplya:2011qq}:
\begin{equation}
dr_*= \frac{d r}{f(r)}.
\end{equation}
The effective potentials for the scalar and electromagnetic fields are:
\begin{equation}\label{scalarpotential}
V_{scal}(r) = f(r)\left(\frac{\ell(\ell+1)}{r^2}+\frac{1}{r}\frac{d f(r)}{dr}\right),
\end{equation}
\begin{equation}\label{empotential}
V_{em}(r) = f(r)\frac{\ell(\ell+1)}{r^2}.
\end{equation}
Thus, $r^{*} \rightarrow -\infty$ corresponds to the black hole event horizon $r_{+}$.

The effective potentials have the form of a positive definite potential barrier with a single maximum, except for the case of $\ell=0$ scalar perturbations, for which the effective potential has a negative gap (see Fig. \ref{fig1}). This means that the stability for this case is not evident and we will test it in the next section.

Quasinormal modes $\omega_{n}$ correspond to solutions of the master wave equation (\ref{klein-Gordon}) with the requirement of the purely outgoing waves at infinity and purely incoming waves at the event horizon (see,  for example, \cite{Konoplya:2011qq,Kokkotas:1999bd}):
\begin{equation}
\Psi_{s} \sim \pm e^{\pm i \omega r^{*}}, \quad r^{*} \rightarrow \pm \infty.
\end{equation}
When the solution is asymptotically de Sitter, purely outgoing waves are required at the de Sitter horizon instead of infinity.

\section{Quasinormal ringing via the WKB and time-domain integration methods}

In order to analyze evolution of perturbations in time-domain we will use the method for integration of the wave equation in time domain, that is, before introduction of the stationary ansatz, at a given point in space \cite{Gundlach:1993tp}.
We will integrate the wavelike equation rewritten in terms of the light-cone variables $u=t-r_*$ and $v=t+r_*$. The appropriate discretization scheme was suggested in \cite{Gundlach:1993tp}:
\begin{eqnarray}\nonumber
\Psi\left(N\right)&=&\Psi\left(W\right)+\Psi\left(E\right)-\Psi\left(S\right)-
\\&&
\Delta^2\frac{V\left(S\right)\left(\Psi\left(W\right)+\Psi\left(E\right)\right)}{8}+{\cal O}\left(\Delta^4\right)\,,
\end{eqnarray}
where the following notation for the points were used:
$N=\left(u+\Delta,v+\Delta\right)$, $W=\left(u+\Delta,v\right)$, $E=\left(u,v+\Delta\right)$ and $S=\left(u,v\right)$. The initial data are given on the null surfaces $u=u_0$ and $v=v_0$. This method was used in a great number of works and proved its efficiency (see for example \cite{Konoplya:2020jgt,Konoplya:2020bxa,Konoplya:2019hml,Churilova:2020bql} and references therein).

In the frequency domain  we will use the WKB method of Will and Schutz \cite{Schutz:1985zz}, which was extended to higher orders in \cite{Iyer:1986np,Konoplya:2003ii,Matyjasek:2017psv} and made even more accurate by the usage of the Pad\'{e} approximants in \cite{Matyjasek:2017psv,Hatsuda:2019eoj}.
The higher-order WKB formula \cite{Konoplya:2019hlu} has the form:
$$ \omega^2=V_0+A_2(\K^2)+A_4(\K^2)+A_6(\K^2)+\ldots- $$
\begin{equation}\nonumber
\imo \K\sqrt{-2V_2}\left(1+A_3(\K^2)+A_5(\K^2)+A_7(\K^2)\ldots\right),
\end{equation}
where $\K$ takes half-integer values. The corrections $A_k(\K^2)$ of order $k$ to the eikonal formula are polynomials of $\K^2$ with rational coefficients and depend on the values of higher derivatives of the potential $V(r)$ in its maximum. In order to increase accuracy of the WKB formula, we will follow Matyjasek and Opala \cite{Matyjasek:2017psv} and use Padé approximants.

As both methods are very well known (\cite{Konoplya:2019hlu,Konoplya:2011qq}), we will not describe them in this paper in more detail, but will simply show that data obtained by  both methods are in a very good agreement.

\begin{table*}
  \centering
\begin{tabular}{|c|c|c|}
  \hline
  $\gamma$ & Time-Domain & WKB \\
  \hline
  0 & $0.585817 - 0.193680 i$, asymptotic \quad tail & $0.585691 - 0.195298 i$, $0.529608 - 0.612039 i$ \\
    \hline
  0.1 & $0.626607 - 0.198156 i$, $0.1086120 - 0.0316408 i$ & $0.626432 - 0.199957 i$, $0.561854 - 0.630399 i$ \\
    \hline
  0.2 & $0.666572 - 0.202136 i$, $0.244780 - 0.057896 i$ & $0.666554 - 0.202376 i$, $0.592440 - 0.642855 i$ \\
    \hline
  0.4 & $0.745891 - 0.201371 i$, $0.447305 - 0.0651241 i$ & $0.745880 - 0.200337 i$, $0.650951 - 0.649154 i$ \\
    \hline
  0.6 & $0.825988 - 0.181181 i$, $0.658562 - 0.0641326 i$ & $0.825624 - 0.188484 i$, $0.711067 - 0.624698 i$ \\
    \hline
  0.8 & $0.906628 - 0.170434 i$, $0.862459 - 0.059459 i$ & $0.907968 - 0.165472 i$, $0.742103 - 0.570844 i$ \\
     \hline
  1 & $1.05984 - 0.124847 i$, $1.05006 - 0.0556556 i$ & $0.994072 - 0.124870 i$, $0.767872 - 0.513711 i$ \\
  \hline
\end{tabular}
  \caption{Quasinormal modes for scalar $s=0$ perturbations for various values of $\gamma$; $r_{+}=1$ $\ell =1$, $k =10^{-4}$.  The second mode in the time domain data represents the second stage of quasinormal ringing induced by the effective dark matter (DM) term.}\label{Table1}
\end{table*}

\begin{table}
  \centering
\begin{tabular}{|c|c|c|}
  \hline
  $\gamma$ & Time-Domain & WKB \\
  \hline
  0 & $0.496520 - 0.184975 i$ &  $0.496520 - 0.184975 i$ \\
    \hline
  0.1 & $0.520591 - 0.193820 i$ & $0.520538 - 0.193898 i$ \\
    \hline
  0.2 & $0.543096 - 0.202234 i$ & $0.543098 - 0.202233 i$ \\
    \hline
  0.4 & $0.584386 - 0.217545 i$ & $0.584385 - 0.217556 i$ \\
    \hline
  0.6 & $0.620287 - 0.231584 i$ & $0.620243 - 0.231503 i$ \\
    \hline
  0.8 & $0.648245 - 0.243741 i$ & $0.648244 - 0.243729  i$ \\
     \hline
  1 & $0.661441 - 0.249998 i$ & $0.661437 - 0.250000  i$ \\
  \hline
\end{tabular}
  \caption{Fundamental ($n=0$) quasinormal modes for electromagnetic $s=1$ perturbations for various values of $\gamma$; $r_{+}=1$, $\ell =1$, $k = 0$. }\label{Table2}
\end{table}

\begin{figure*}
\includegraphics[angle=0.0,width=0.5\linewidth]{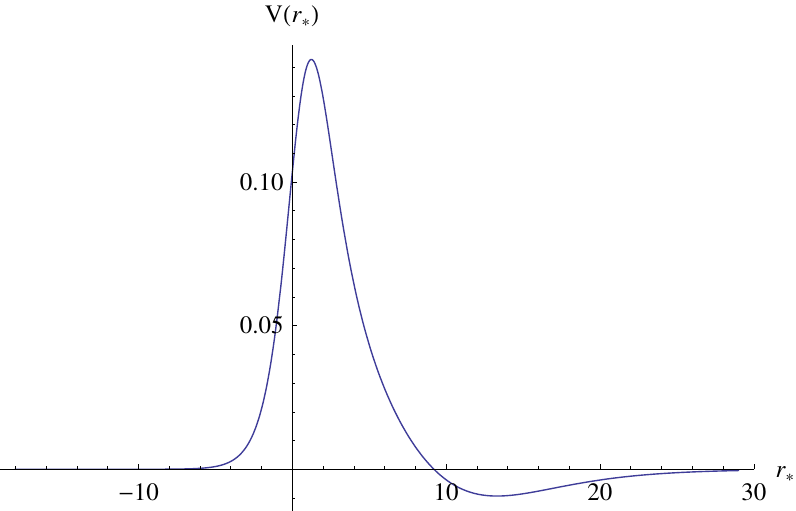}\includegraphics[angle=0.0,width=0.5\linewidth]{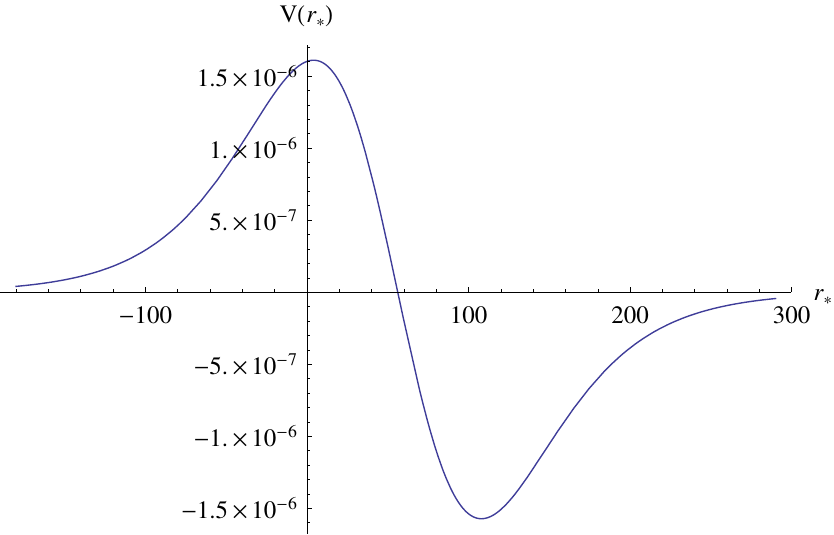}
\caption{Effective potentials for scalar perturbations $\gamma=0.2$, $k=10^{-2}$ (left) and $k=0.39$ (right), $\ell=0$, $r_{+}=1$. }
\label{fig1}
\end{figure*}

\begin{figure*}
\includegraphics[angle=0.0,width=0.7\linewidth]{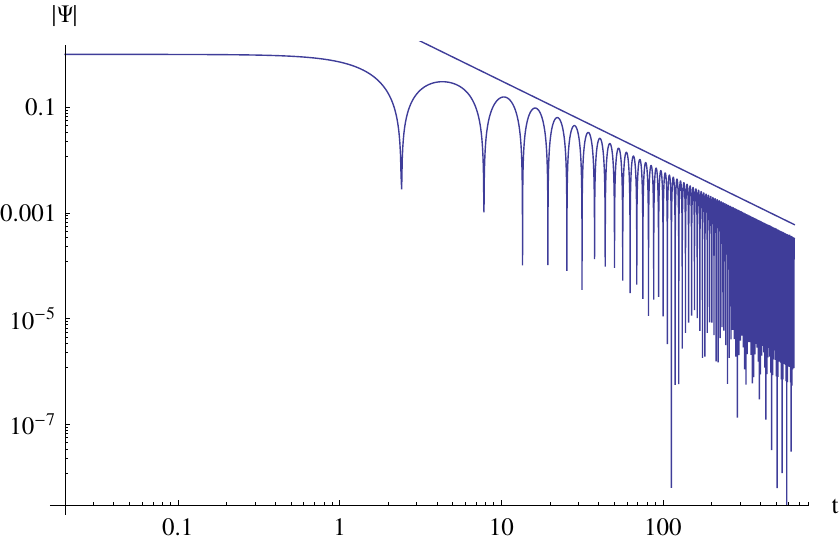}
\caption{Logarithmic plot of the time-domain evolution of scalar perturbations $\gamma=0.5$, $k=0$, $\ell=0$ (left); $r_{+}=1$. }
\label{fig2}
\end{figure*}

\begin{figure*}
\includegraphics[angle=0.0,width=0.7\linewidth]{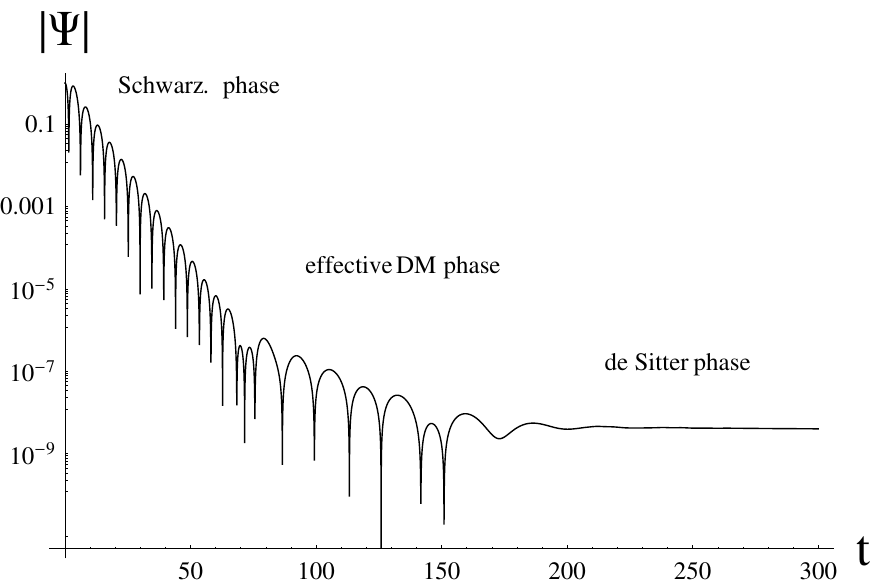}
\caption{Semilogarithmic plot of the time-domain evolution of scalar perturbations $\gamma=0.2$, $k=10^{-4}$,  $\ell=1$; $r_{+}=1$. }
\label{fig3}
\end{figure*}

\begin{figure*}
\includegraphics[angle=0.0,width=0.5\linewidth]{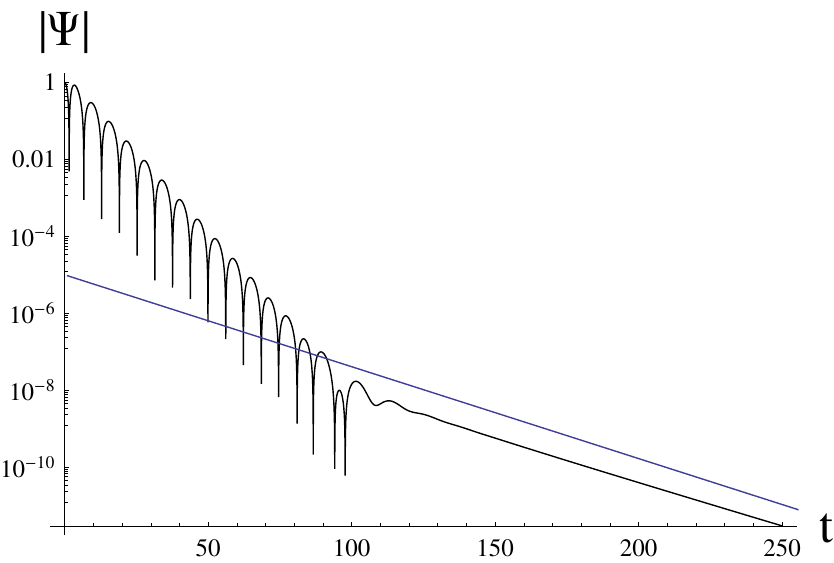}\includegraphics[angle=0.0,width=0.5\linewidth]{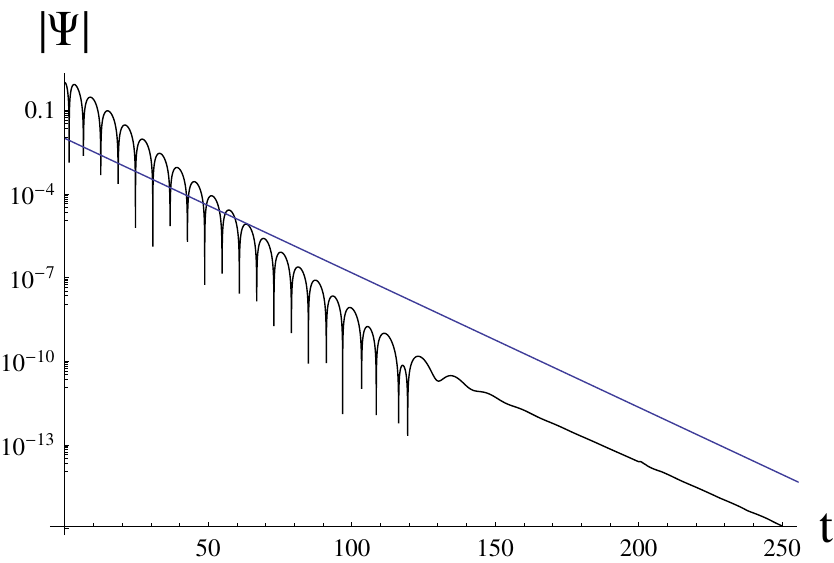}
\caption{Semilogarithmic plot of the time-domain evolution of electromagnetic perturbations for $\gamma=0.05$ (left) and $\gamma=0.1$ (right); $\ell=1$, $r_{+}=1$, $k=0$. The straight lines represent the exponential decays $ \sim e^{-0.055 t}$ (left) and $ \sim e^{-0.11 t}$ (right). }
\label{fig4}
\end{figure*}

First of all, the time-domain integration shows that once the effective cosmological constant is zero, which corresponds to $k=0$, we have the following decay law of a scalar field at asymptotically late times (see an example on Fig. \ref{fig2} for $\ell=0$ perturbations):
\begin{equation}\label{effectively-massive}
|\Psi_s| \sim t^{-3/2} \sin (A t), \quad \ell =0, 1, 2, \ldots, \quad k=0.
\end{equation}
Here the constant $A$ depends on the black hole parameters. The enveloping decay law $\sim t^{-3/2}$ is the same for all multipole numbers $\ell$ and black hole parameters $\gamma$ and $\beta$.
The electromagnetic perturbations decay according to the exponential law at asymptotic times even when the effective cosmological constant is zero:
 \begin{equation}
|\Psi_{el}| \sim e^{-C t}, \quad \ell = 1, 2, \ldots, \quad k=0,
\end{equation}
where the factor $C$ depends on $\gamma$ and $\ell$ (see, as an example, Fig. \ref{fig4}), for example,
 \begin{equation}
C \approx 1.1 \gamma, \quad \ell =1.
\end{equation}

When $k$ is not zero, we have the following asymptotic decay law for a massless scalar field:
\begin{equation}
|\Psi_{s}| \sim |\Psi_{0}| + |\Psi_{1}|  e^{-p_{s} t}, \quad \ell=0, 1, \ldots, k \neq 0,
\end{equation}
where $p_s$ depends on the black hole parameters (see, for instance, Fig. \ref{fig3}).
The electromagnetic field decays in the presence of the effective de Sitter term according to the following law:
\begin{equation}
|\Psi_{el}| \sim e^{- C t}, \quad \ell = 1, 2, \ldots, k \neq 0,
\end{equation}
where the constant $C$ is the same as in the case $k=0$ above at least for relatively small and moderate values of $k$.

When both $k$ and $\gamma$ are nonzero, the evolution of perturbations of the scalar field consists of the three stages:
\begin{itemize}
\item The quasinormal ringing with the perturbative Schwarzschild frequency, that with the frequency which slowly changes from its Schwarzschild value when $k$ and $\gamma$ are turned on.
\item The second stage of quasinormal ringing with another dominant frequency which is slower damped than the Schwarzschild one (see Table I). When $\gamma$ goes to zero, this long-lived frequency goes over into the purely imaginary "mode" representing the asymptotic de Sitter tail \cite{Brady270}. When $k$ vanishes, these modes are reduced to oscillatory tails enveloped by the power law decay given by the universal law (Eq. \ref{effectively-massive}).
\item The asymptotic exponential tails. Here, unlike the well-known de Sitter tails \cite{Brady270}, the constant $|\Psi_{0}|$ is added to the exponentially decaying term not only for zero multipole, but also for higher multipoles $\ell$.
\end{itemize}

Special attention must be paid to the case $\ell=0$ and the near extremal values of $k$, because the effective potential has a deep negative gap (see Fig. \ref{fig1}). There are a number of examples when such negative gaps lead to the unbounded growth of the perturbations, signifying the dynamical instability \cite{Konoplya:2014lha,Zhu:2014sya,Konoplya:2008au,Konoplya:2013sba}. Therefore, the stability must be checked numerically for this case  by the time-domain integration which includes contribution of all the overtones. Any numerical method in the frequency domain will not exclude the possibility of missing the mode leading to instability. A recent study of this near extremal case \cite{Momennia:2019cfd} simply ignores the $\ell=0$ modes and claims the stability based on the positiveness of the effective potential for $\ell =1, 2,...$. Time-domain profiles obtained here for near extremal cases as well show that the scalar field is apparently stable even for $\ell=0$, because the wave function decays in time.

It is worth mentioning that the dominant quasinormal modes extracted from the time domain profiles by the Prony method are in a very good agreement with those obtained via the 7th order WKB method with further usage of the Padé approximants as prescribed in \cite{Matyjasek:2017psv}. This can be seen in data presented in Tables I and II. The choice of the Padé approximants was such that the known accurate quasinormal modes of Schwarzschild black hole is reproduced with the best accuracy, which corresponds to $\tilde{m} =7$, where $\tilde{m}$ is defined in \cite{Konoplya:2019hlu}.

\section{Discussion}

In the present paper we considered the evolution of scalar and electromagnetic perturbations in the vicinity of the  Mannheim-Kazanas black hole solution \cite{MK} in the conformal Weyl gravity. We found a number of peculiarities, which were omitted in previous studies of quasinormal modes in this theory in the frequency domain \cite{Momennia:2019cfd,Mehrab-Momennia,Momennia:2019edt}:

\begin{itemize}
\item We have shown that $\ell=0$ scalar field perturbations of the nearly extremal black holes are governed by the effective potential which has a deep negative gap, and, that, nevertheless, time-domain profiles are decaying, which points to the stability of the scalar field.
\item When the effective de Sitter term vanishes ($k=0$), the asymptotic tails of the massless scalar fields are not power-law, as it happens for the Schwarzschild case  \cite{Price267,Bicak268}, but the oscillatory ones with a power-law enveloping of oscillations, like it happens for massive fields in the background of asymptotically flat black holes   \cite{Burko:2004jn,274Koyama}.
\item When the effective de Sitter term is turned on, this oscillatory tail goes over into the exponential quasinormal ringing dominated by an essentially non-Schwarzschildian, longer-lived, frequency.
\item The asymptotic decay law at $t \rightarrow \infty$ for the scalar field is, then, exponential, so that the whole evolution of the signal consists of the three stages: the first stage of  quasinormal ringing at the Schwarzschild-like frequency, the second stage of quasinormal ringing at the long-lived non-Schwarzschild frequency and  the exponential tail.
\item The asymptotic tails for electromagnetic field are exponential even when the effective cosmological term is tuned off ($k=0$).

\end{itemize}

Our work could be extended in a number of ways. First of all, one could make an analytical derivation of the asymptotic behavior via the analysis of the asymptotic behavior of the wave equations in a similar fashion with \cite{Ching266}. Then the same analysis could be done for the Dirac perturbations describing the neutrino field. In the latter case we would face the stability problem as well \cite{LopezOrtega:2012hx,Konoplya:2020zso}. A much more complicated problem would be the case of gravitational quasinormal modes which has not been performed in \cite{Momennia:2019edt}, because the Weyl equations were not perturbed and, instead, perturbation of the same class of metrics, but in an essentially different theory was considered.

\acknowledgments{The author acknowledges 19-03950S GAČR grant and the RUDN University Program 5-100.}


\begin{thebibliography}{80}
\bibitem{Konoplya:2011qq}
  R.~A.~Konoplya and A.~Zhidenko,
  Rev.\ Mod.\ Phys.\  {\bf 83}, 793 (2011)
  \href{https://arxiv.org/abs/1102.4014}{[arXiv:1102.4014 [gr-qc]]}.
\bibitem{Kokkotas:1999bd}
  K.~D.~Kokkotas and B.~G.~Schmidt,
  Living Rev.\ Rel.\  {\bf 2} (1999) 2
  [gr-qc/9909058].
\bibitem{Berti:2009kk}
  E.~Berti, V.~Cardoso and A.~O.~Starinets,
  Class.\ Quant.\ Grav.\  {\bf 26}, 163001 (2009)
  [arXiv:0905.2975 [gr-qc]].
\bibitem{LIGO}
  B.~P.~Abbott {\it et al.}
  Phys.\ Rev.\ Lett.\  {\bf 116}, no. 6, 061102 (2016)
  [arXiv:1602.03837 [gr-qc]];
\bibitem{Price267} Price, R. H., Phys. Rev. D 5, 2419 (1972); Phys. Rev. D5, 2439 (1972).
\bibitem{Bicak268}  Bicak, J., Gen. Relativ. Gravit. 3, 331 (1972).
\bibitem{Ching266} Ching, E. S. C., Leung, P. T., Suen, W. M. and Young, K., Phys. Rev. D 52, 2118 (1995) [arXiv:gr-qc/9507035].
\bibitem{Ching:1994bd}
  E.~S.~C.~Ching, P.~T.~Leung, W.~M.~Suen and K.~Young,
  Phys.\ Rev.\ Lett.\  {\bf 74}, 2414 (1995)
  [gr-qc/9410044].
\bibitem{Burko:2004jn}
  L.~M.~Burko and G.~Khanna,
  Phys.\ Rev.\ D {\bf 70}, 044018 (2004)
  [gr-qc/0403018].
\bibitem{Brady270} Brady, P. R., Chambers, C. M., Laarakkers, W. G., and Poisson, E., Phys. Rev. D60, 064003 (1999).
\bibitem{Churilova:2019qph}
  M.~S.~Churilova, R.~A.~Konoplya and A.~Zhidenko,
  Phys.\ Lett.\ B {\bf 802}, 135207 (2020)
  [arXiv:1911.05246 [gr-qc]].
\bibitem{Konoplya:2006gq}
  R.~A.~Konoplya, A.~Zhidenko and C.~Molina,
  Phys.\ Rev.\ D {\bf 75}, 084004 (2007)
  [gr-qc/0602047].
\bibitem{Rogatko:2008ut}
  M.~Rogatko and A.~Szyplowska,
  Gen.\ Rel.\ Grav.\  {\bf 41}, 1611 (2009)
  [arXiv:0812.1644 [hep-th]].
\bibitem{Gibbons:2008gg}
  G.~W.~Gibbons, M.~Rogatko and A.~Szyplowska,
  Phys.\ Rev.\ D {\bf 77}, 064024 (2008)
  [arXiv:0802.3259 [hep-th]].
\bibitem{Gibbons:2008rs}
  G.~W.~Gibbons and M.~Rogatko,
  Phys.\ Rev.\ D {\bf 77}, 044034 (2008)
  [arXiv:0801.3130 [hep-th]].
\bibitem{Rogatko:2007zz}
  M.~Rogatko and A.~Szyplowska,
  Phys.\ Rev.\ D {\bf 76}, 044010 (2007).
\bibitem{Moderski:2001gt}
  R.~Moderski and M.~Rogatko,
  Phys.\ Rev.\ D {\bf 63}, 084014 (2001).
\bibitem{275Jing}  Jing, J., Phys. Rev. D 72, 027501 (2005) [arXiv:grqc/0408090].
\bibitem{Brady:1996za}
  P.~R.~Brady, C.~M.~Chambers, W.~Krivan and P.~Laguna,
  Phys.\ Rev.\ D {\bf 55}, 7538 (1997)
  [gr-qc/9611056].
\bibitem{274Koyama} Koyama, H. and Tomimatsu, A., Phys. Rev. D 63,
064032 (2001) [arXiv:gr-qc/0012022]; Phys. Rev. D 64,
044014 (2001) [arXiv:gr-qc/0103086]; Phys. Rev. D 65,
084031 (2002) [arXiv:gr-qc/0112075].
\bibitem{Konoplya:2005et}
  R.~A.~Konoplya and C.~Molina,
  Phys.\ Rev.\ D {\bf 71}, 124009 (2005)
  [gr-qc/0504139].
\bibitem{Konoplya:2013rxa}
  R.~A.~Konoplya and A.~Zhidenko,
  Phys.\ Rev.\ D {\bf 88}, 024054 (2013)
  [arXiv:1307.1812 [gr-qc]].
\bibitem{Cardoso:2017cqb}
  V.~Cardoso and P.~Pani,
  Nature Astron.\  {\bf 1}, no. 9, 586 (2017)
  [arXiv:1709.01525 [gr-qc]].
\bibitem{MK}
P. D. Mannheim, D. Kazanas,
Astrophys. J. {\bf 342}, 635 (1989).
\bibitem{Bach}
R. Bach,
Math. Zeit. {\bf 9}, 110 (1921).
\bibitem{Riegert}
R. J. Riegert, Phys. Rev. Lett. {\bf 53}, 315 (1984).
\bibitem{Molina:2003dc}
  C.~Molina, D.~Giugno, E.~Abdalla and A.~Saa,
  Phys.\ Rev.\ D {\bf 69}, 104013 (2004)
  [gr-qc/0309079].
\bibitem{Islam:2018ymd}
  T.~Islam,
  Mon.\ Not.\ Roy.\ Astron.\ Soc.\  {\bf 488}, no. 4, 5390 (2019)
  [arXiv:1811.00065 [gr-qc]].
\bibitem{Dutta:2018oaj}
  K.~Dutta and T.~Islam,
  Phys.\ Rev.\ D {\bf 98}, no. 12, 124012 (2018)
  [arXiv:1808.06923 [gr-qc]].
\bibitem{Christodoulou:2018xxw}
  D.~M.~Christodoulou and D.~Kazanas,
  Mon.\ Not.\ Roy.\ Astron.\ Soc.\  {\bf 479}, no. 1, L143 (2018)
  [arXiv:1806.09778 [gr-qc]].
\bibitem{Takizawa:2020dja}
  K.~Takizawa, T.~Ono and H.~Asada,
  Phys.\ Rev.\ D {\bf 102}, no. 6, 064060 (2020)
  [arXiv:2006.00682 [gr-qc]].
\bibitem{Kasikci:2018mtg}
  O.~Kaşıkçı and C.~Deliduman,
  Phys.\ Rev.\ D {\bf 100}, no. 2, 024019 (2019)
  [arXiv:1812.01076 [gr-qc]].
\bibitem{Fathi:2020sey}
  M.~Fathi, M.~Kariminezhad, M.~Olivares and J.~R.~Villanueva,
  Eur.\ Phys.\ J.\ C {\bf 80}, no. 5, 377 (2020)
  [arXiv:2009.03399 [gr-qc]].
\bibitem{Turner:2020gxo}
  G.~E.~Turner and K.~Horne,
  Class.\ Quant.\ Grav.\  {\bf 37}, no. 9, 095012 (2020).
\bibitem{Li:2020wvn}
  Z.~Li, G.~Zhang and A.~Övgün,
  Phys.\ Rev.\ D {\bf 101}, no. 12, 124058 (2020)
  [arXiv:2006.13047 [gr-qc]].
\bibitem{Fathi:2019jgd}
  M.~Fathi, M.~Olivares and J.~R.~Villanueva,
  Eur.\ Phys.\ J.\ C {\bf 80}, no. 1, 51 (2020)
  [arXiv:1910.12811 [gr-qc]].
\bibitem{Fathi:2020sfw}
  M.~Fathi, M.~Olivares and J.~R.~Villanueva,
  arXiv:2009.03404 [gr-qc].
\bibitem{Lanteri:2020trb}
  D.~Lanteri, S.~S.~Wan, A.~Iorio and P.~Castorina,
  arXiv:2009.14087 [hep-th].
\bibitem{Xu:2018liy}
  H.~Xu and M.~H.~Yung,
  Phys.\ Lett.\ B {\bf 793}, 97 (2019)
  [arXiv:1811.07309 [gr-qc]].
\bibitem{Momennia:2019cfd}
  M.~Momennia and S.~H.~Hendi,
  Phys.\ Rev.\ D {\bf 99}, no. 12, 124025 (2019)
  [arXiv:1905.12290 [gr-qc]].
\bibitem{Mehrab-Momennia} M. Momennia, S. H. Hendi, F. S. Bidgoli,
Phys. Lett. B, Vol. 813, 136028 (2021)
\bibitem{Momennia:2019edt}
  M.~Momennia and S.~H.~Hendi,
  Eur.\ Phys.\ J.\ C {\bf 80}, no. 6, 505 (2020)
  [arXiv:1910.00428 [gr-qc]].
\bibitem{Sharif:2020icx}
  M.~Sharif and Z.~Akhtar,
  Phys.\ Dark Univ.\  {\bf 29}, 100589 (2020)
  [arXiv:2005.09430 [gr-qc]].
\bibitem{Gundlach:1993tp}
  C.~Gundlach, R.~H.~Price and J.~Pullin,
  Phys.\ Rev.\ D {\bf 49}, 883 (1994)
  [gr-qc/9307009].
\bibitem{Konoplya:2020jgt}
  R.~A.~Konoplya, A.~F.~Zinhailo and Z.~Stuchlik,
  Phys.\ Rev.\ D {\bf 102}, no. 4, 044023 (2020)
  [arXiv:2006.10462 [gr-qc]].
\bibitem{Konoplya:2020bxa}
  R.~A.~Konoplya and A.~F.~Zinhailo,
  Eur.\ Phys.\ J.\ C {\bf 80}, no. 11, 1049 (2020)
  [arXiv:2003.01188 [gr-qc]].
\bibitem{Konoplya:2019hml}
  R.~A.~Konoplya, A.~F.~Zinhailo and Z.~Stuchlík,
  Phys.\ Rev.\ D {\bf 99}, no. 12, 124042 (2019)
  [arXiv:1903.03483 [gr-qc]].
\bibitem{Churilova:2020bql}
  M.~S.~Churilova,
  Phys.\ Rev.\ D {\bf 102}, no. 2, 024076 (2020)
  [arXiv:2002.03450 [gr-qc]].
\bibitem{Schutz:1985zz}
  B.~F.~Schutz and C.~M.~Will,
  Astrophys.\ J.\  {\bf 291}, L33 (1985).
\bibitem{Iyer:1986np}
  S.~Iyer and C.~M.~Will,
  Phys.\ Rev.\ D {\bf 35}, 3621 (1987).
\bibitem{Konoplya:2003ii}
  R.~A.~Konoplya,
  Phys.\ Rev.\ D {\bf 68}, 024018 (2003)
  \href{https://arxiv.org/abs/gr-qc/0303052}{[gr-qc/0303052]}.
\bibitem{Matyjasek:2017psv}
  J.~Matyjasek and M.~Opala,
  Phys.\ Rev.\ D {\bf 96}, no. 2, 024011 (2017)
  \href{https://arxiv.org/abs/1704.00361}{[arXiv:1704.00361 [gr-qc]]}.
\bibitem{Hatsuda:2019eoj}
  Y.~Hatsuda,
  arXiv:1906.07232 [gr-qc].
\bibitem{Konoplya:2019hlu}
  R.~A.~Konoplya, A.~Zhidenko and A.~F.~Zinhailo,
  Class.\ Quant.\ Grav.\  {\bf 36}, 155002 (2019)
  \href{https://arxiv.org/abs/1904.10333}{[arXiv:1904.10333 [gr-qc]]}.
\bibitem{Konoplya:2014lha}
  R.~A.~Konoplya and A.~Zhidenko,
  Phys.\ Rev.\ D {\bf 90}, no. 6, 064048 (2014)
  [arXiv:1406.0019 [hep-th]].
\bibitem{Zhu:2014sya}
  Z.~Zhu, S.~J.~Zhang, C.~E.~Pellicer, B.~Wang and E.~Abdalla,
  Phys.\ Rev.\ D {\bf 90}, no. 4, 044042 (2014)
  Addendum: [Phys.\ Rev.\ D {\bf 90}, no. 4, 049904 (2014)]
  [arXiv:1405.4931 [hep-th]].
\bibitem{Konoplya:2008au}
  R.~A.~Konoplya and A.~Zhidenko,
  Phys.\ Rev.\ Lett.\  {\bf 103}, 161101 (2009)
  [arXiv:0809.2822 [hep-th]].
\bibitem{Konoplya:2013sba}
  R.~A.~Konoplya and A.~Zhidenko,
  Phys.\ Rev.\ D {\bf 89}, no. 2, 024011 (2014)
  [arXiv:1309.7667 [hep-th]].
\bibitem{LopezOrtega:2012hx}
  A.~Lopez-Ortega,
  Int.\ J.\ Mod.\ Phys.\ D {\bf 21}, 1250092 (2012)
  [arXiv:1211.1801 [gr-qc]].
\bibitem{Konoplya:2020zso}
  R.~A.~Konoplya and M.~S.~Churilova,
  arXiv:2004.05879 [gr-qc].











\end{thebibliography}
\end{document}